\documentclass{PoS}
\usepackage{wrapfig}
\usepackage{url}

\title{New results on fluctuations and correlations from the NA61/SHINE experiment at the CERN SPS}

\ShortTitle{Fluctuations and correlations from NA61/SHINE}

\author{\speaker{Katarzyna Grebieszkow}\thanks{This work was partially supported by the National Science Centre, Poland (grant 2015/18/M/ST2/00125).} \hspace{0.1cm} for the  NA61/SHINE Collaboration \\
        Warsaw University of Technology \\
        E-mail: \email{kperl@if.pw.edu.pl}}

\abstract{The exploration of the QCD phase diagram is the most important task of present heavy ion experiments. In particular, we want to study the phase transition from hadronic to partonic matter and look for the critical point (CP) of strongly interacting matter. Fluctuations and correlations in kinematic characteristics and particle yields may help to locate the CP (in analogy to enlarged fluctuations due to critical opalescence close to a CP in a liquid/gas transition). The strong interactions program of the NA61/SHINE experiment may allow to discover or rule out the existence of the CP in the Super Proton Synchrotron energy domain. For this purpose we perform a two-dimensional scan 
by varying the energy ($5.1 < \sqrt{s_{NN}} < 16.8/17.3$ GeV) and the system size (p+p, p+Pb, Be+Be, Ar+Sc, Xe+La, Pb+Pb) of the collisions.

In this report new NA61/SHINE results on fluctuations and correlations in p+p, Be+Be, and Ar+Sc collisions are presented. In particular, results on transverse momentum and multiplicity fluctuations, as well as higher order moments of net-charge fluctuations are discussed. The NA61/SHINE data are compared to predictions of string hadronic models and to NA49 results.}

\FullConference{The European Physical Society Conference on High Energy Physics\\
          5-12 July\\
          Venice, Italy}

\begin{document}

\section{Introduction and motivation}

There are two important reasons for studying fluctuations and correlations. First, they may serve as a signature of the onset of deconfinement because close to the phase transition the Equation of State changes rapidly. This can impact the energy dependence of fluctuations. Second, they can help to locate the critical point (CP) of strongly interacting matter. This is in analogy to critical opalescence - enlarged fluctuations close to the CP in a liquid/gas transition. For strongly interacting matter a maximum of the CP signal is expected when freeze-out happens near the CP. 

Within the NA61/SHINE strong interactions program we perform a comprehensive scan in the whole Super Proton Synchrotron (SPS) energy range (beam momenta 13$A$-150/158$A$ GeV/c; $\sqrt{s_{NN}}$ = 5.1-16.8/17.3 GeV) with light, intermediate mass and heavy nuclei. So far complete energy scans (six energies) of p+p, Be+Be and Ar+Sc collisions have been recorded, as well as selected energies for Pb+Pb and p+Pb reactions. This year energy scan of p+Pb will be continued and Xe+La collisions will be recorded for the first time. Future large statistics Pb+Pb interactions are planned to be recorded with a new Vertex Detector for open charm and multi-strange particle measurement.      

The experimental strategy of looking for the critical point is based on a search for non-monotonic behavior of CP signatures such as fluctuations of transverse momentum, multiplicity, intermittency, etc. when changing collisions energy and/or size of colliding nuclei.

\section{Transverse momentum and multiplicity fluctuations in p+p, Be+Be and Ar+Sc}

NA61/SHINE uses the {\it strongly intensive} measures $\Delta[P_T, N]$ and $\Sigma[P_T, N]$ to study transverse momentum and multiplicity fluctuations \cite{Aduszkiewicz:2015jna}. In the Wounded Nucleon Model (WNM) they depend neither on the number of wounded nucleons ($W$) nor on fluctuations of $W$. In the Grand Canonical Ensemble they do not depend on volume and volume fluctuations. Moreover, $\Delta[P_T, N]$ and $\Sigma[P_T, N]$ have two reference values, namely they are equal to zero in case of no fluctuations and one in case of independent particle production. 

The NA61/SHINE acceptance used in this analysis is described in Ref.~\cite{Aduszkiewicz:2015jna}. Additionally, due to poor azimuthal angle acceptance and stronger electron contamination at backward rapidities, we limited the rapidity acceptance to $0 < y_{\pi} < y_{beam}$, where $y_{\pi}$ is the rapidity calculated assuming pion mass, and $y_{beam}$ is the beam rapidity. 

\begin{figure}[h]
\centering
\includegraphics[width=0.3\textwidth]{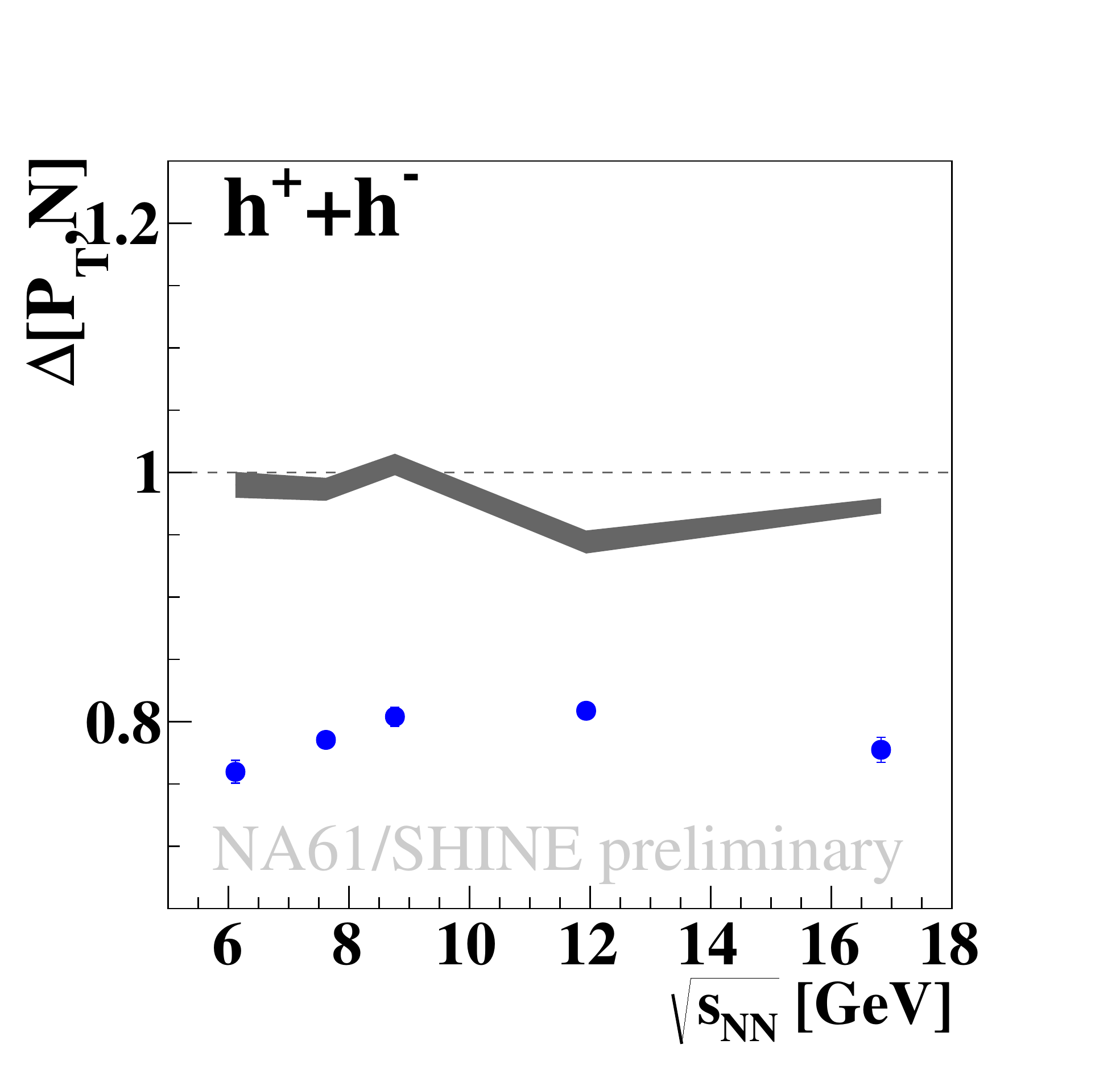}
\includegraphics[width=0.3\textwidth]{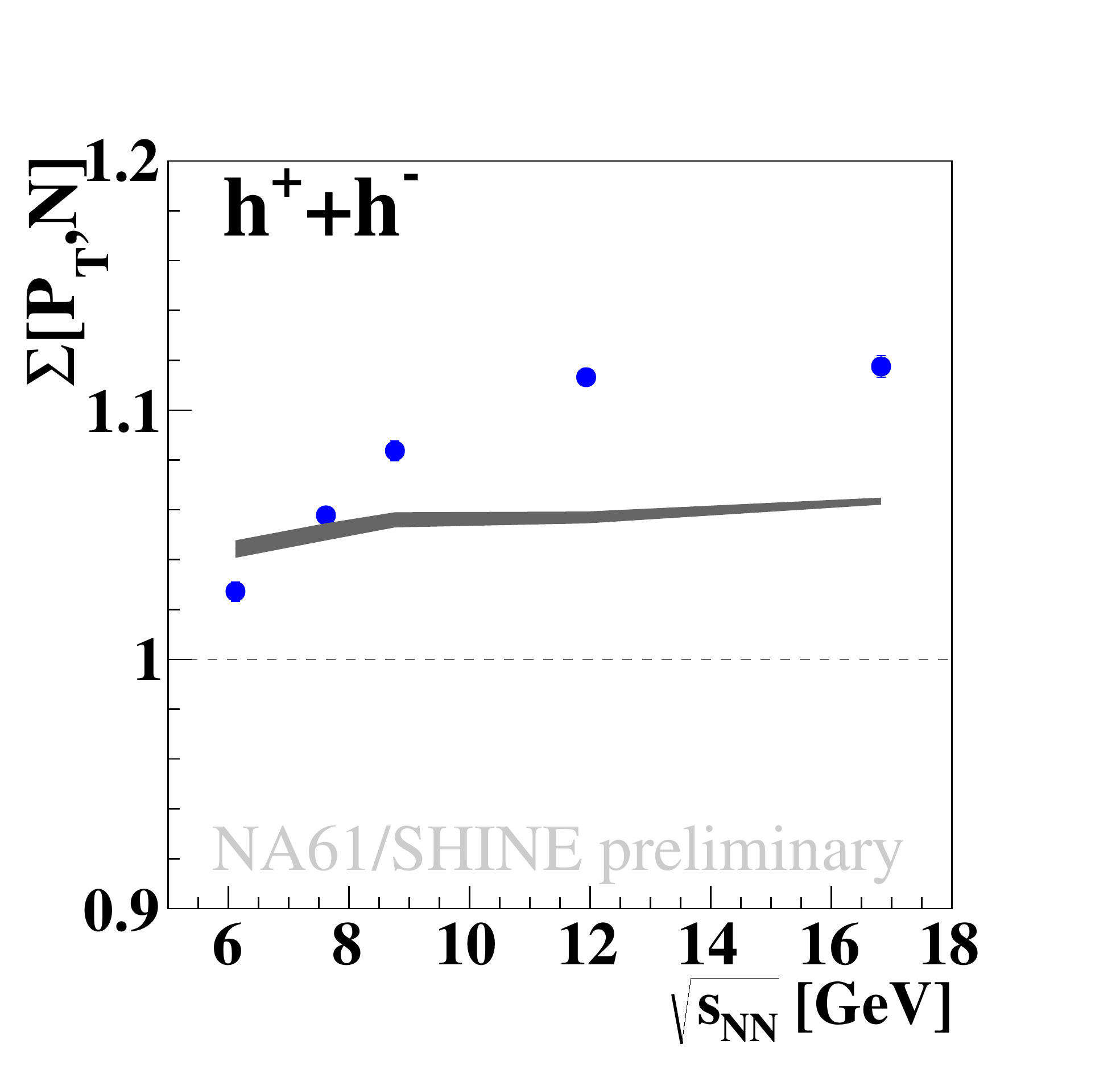}
\vspace{-0.2cm}
\caption[]{\footnotesize $\Delta[P_T, N]$ and $\Sigma[P_T, N]$ measures of transverse momentum and multiplicity fluctuations obtained by NA61/SHINE for 0-5\% most central Ar+Sc collisions at forward rapidity, $0< y_{\pi}< y_{beam}$, and with $p_T<1.5$ GeV/c for all charged hadrons ($h^{+}+h^{-}$). Only statistical uncertainties are shown. Shaded bands show the EPOS 1.99 model.}
\label{evgeny_arsc_epos}
\end{figure}

Figure \ref{evgeny_arsc_epos} shows the energy dependence of $\Delta[P_T, N]$ and $\Sigma[P_T, N]$ measured for the 0-5\% most central Ar+Sc collisions; centrality is defined by the energy measured in the projectile spectator region, with the smallest values corresponding to the most violent (central) collisions. There is no prominent non-monotonic behavior that can be attributed to a CP. The values of $\Delta[P_T, N]$ smaller than 1 and $\Sigma[P_T, N]$ higher than 1 may be due to Bose-Einstein statistics
and/or anti-correlation between event transverse momentum and multiplicity (see Ref.~\cite{Gorenstein:2013nea} and references therein). The NA61/SHINE results are not reproduced by the EPOS 1.99 model.  

\begin{figure}[h]
\centering
\includegraphics[width=0.3\textwidth]{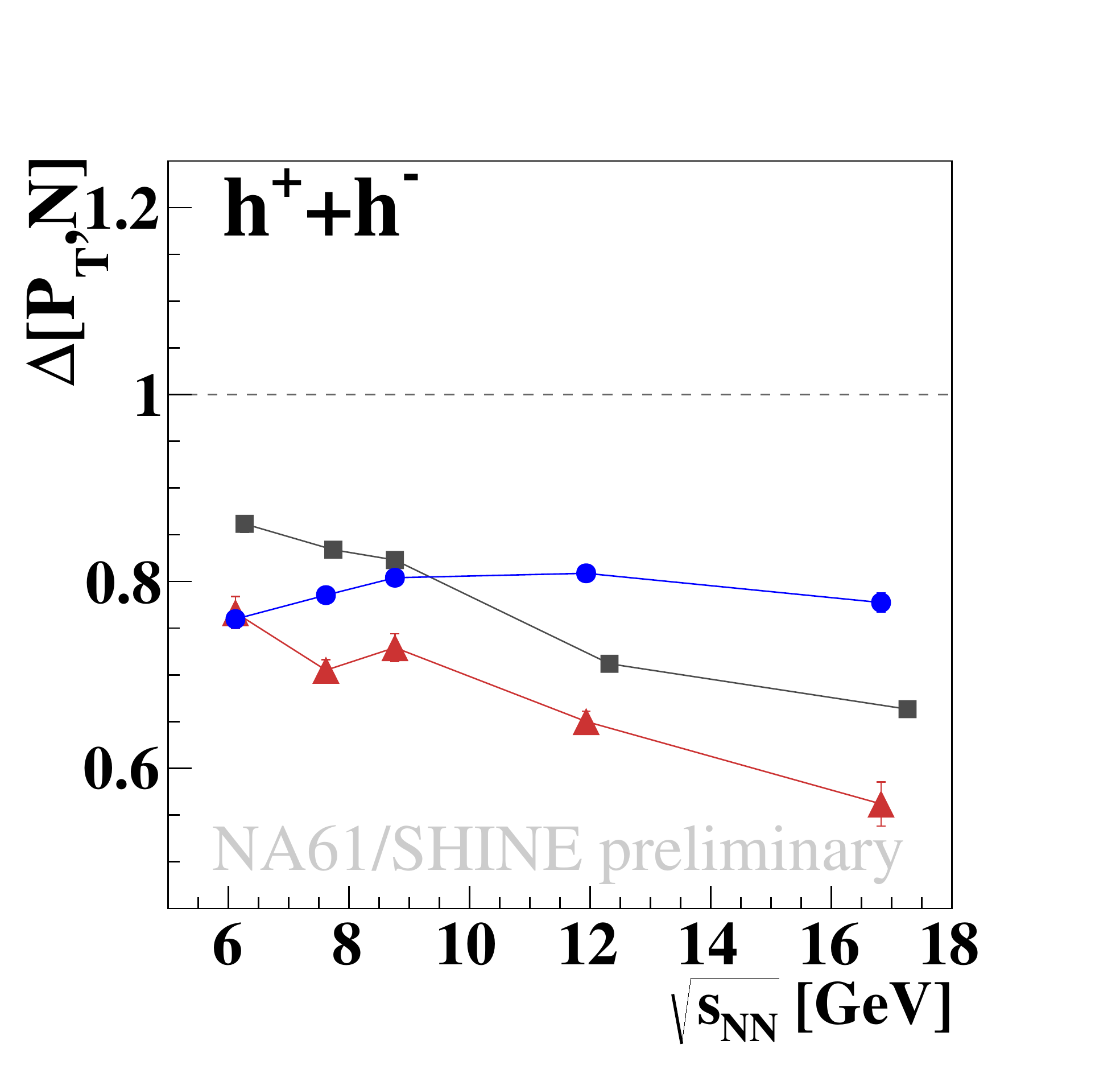}    
\includegraphics[width=0.3\textwidth]{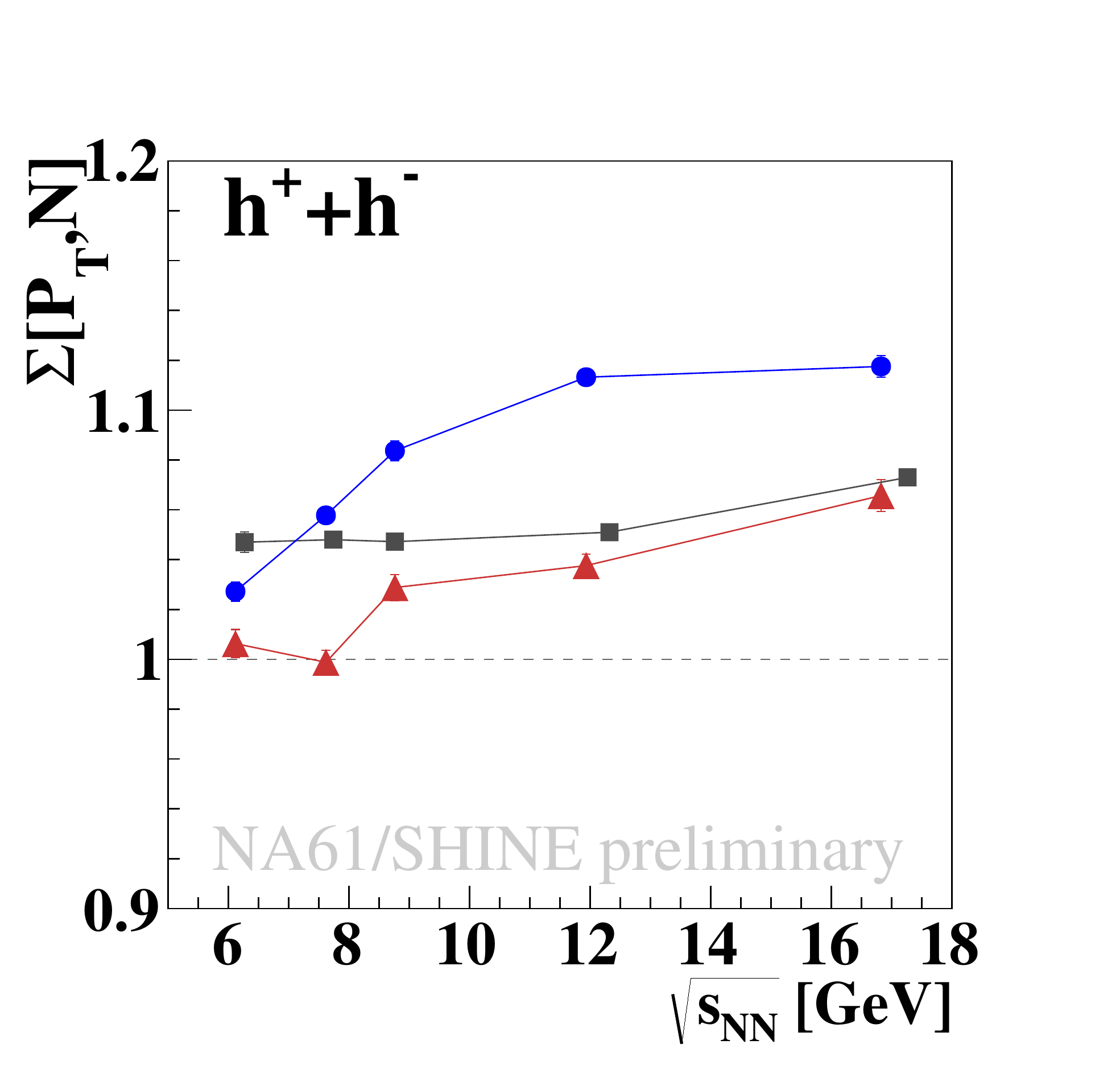} \\
\includegraphics[width=0.35\textwidth]{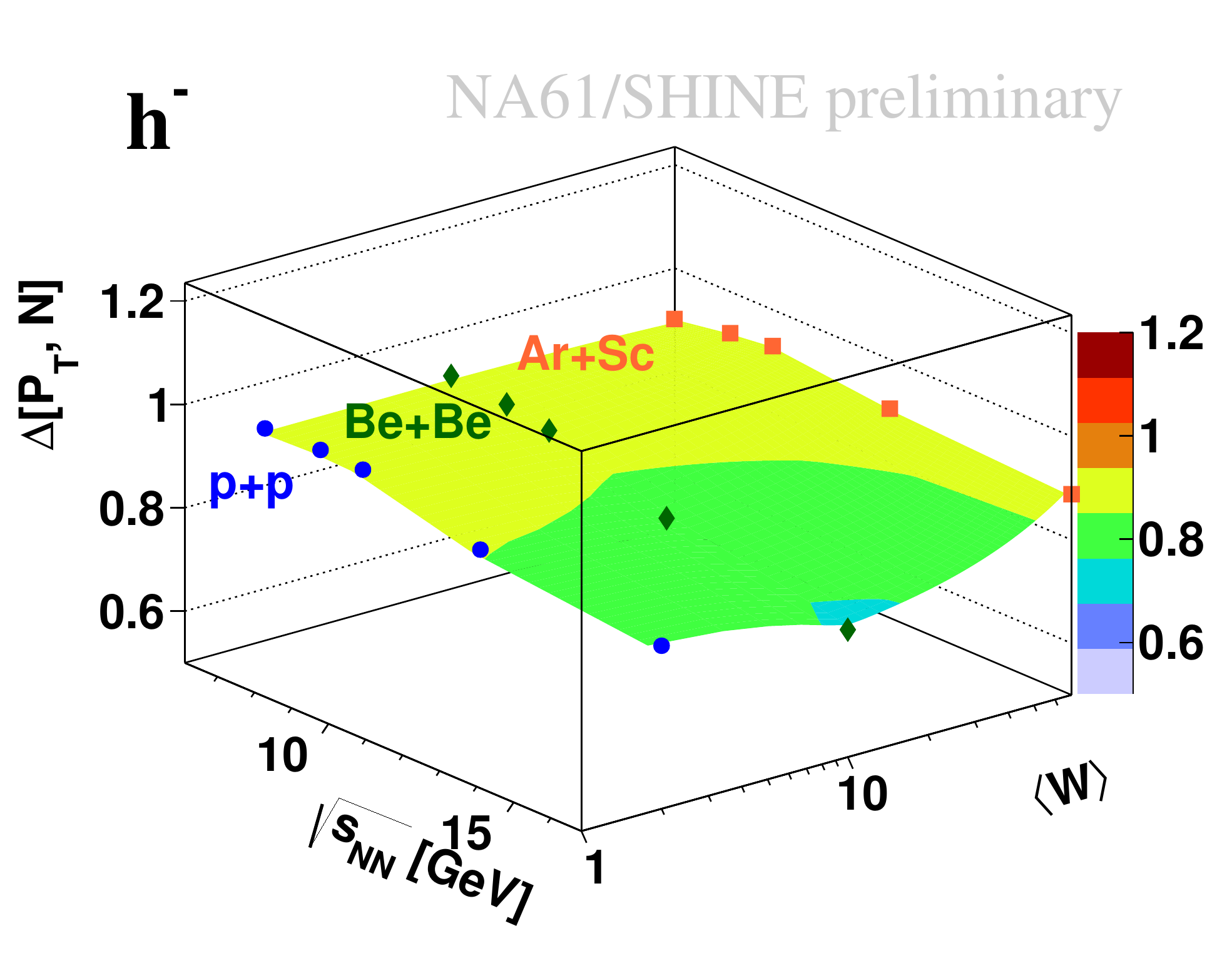}
\includegraphics[width=0.35\textwidth]{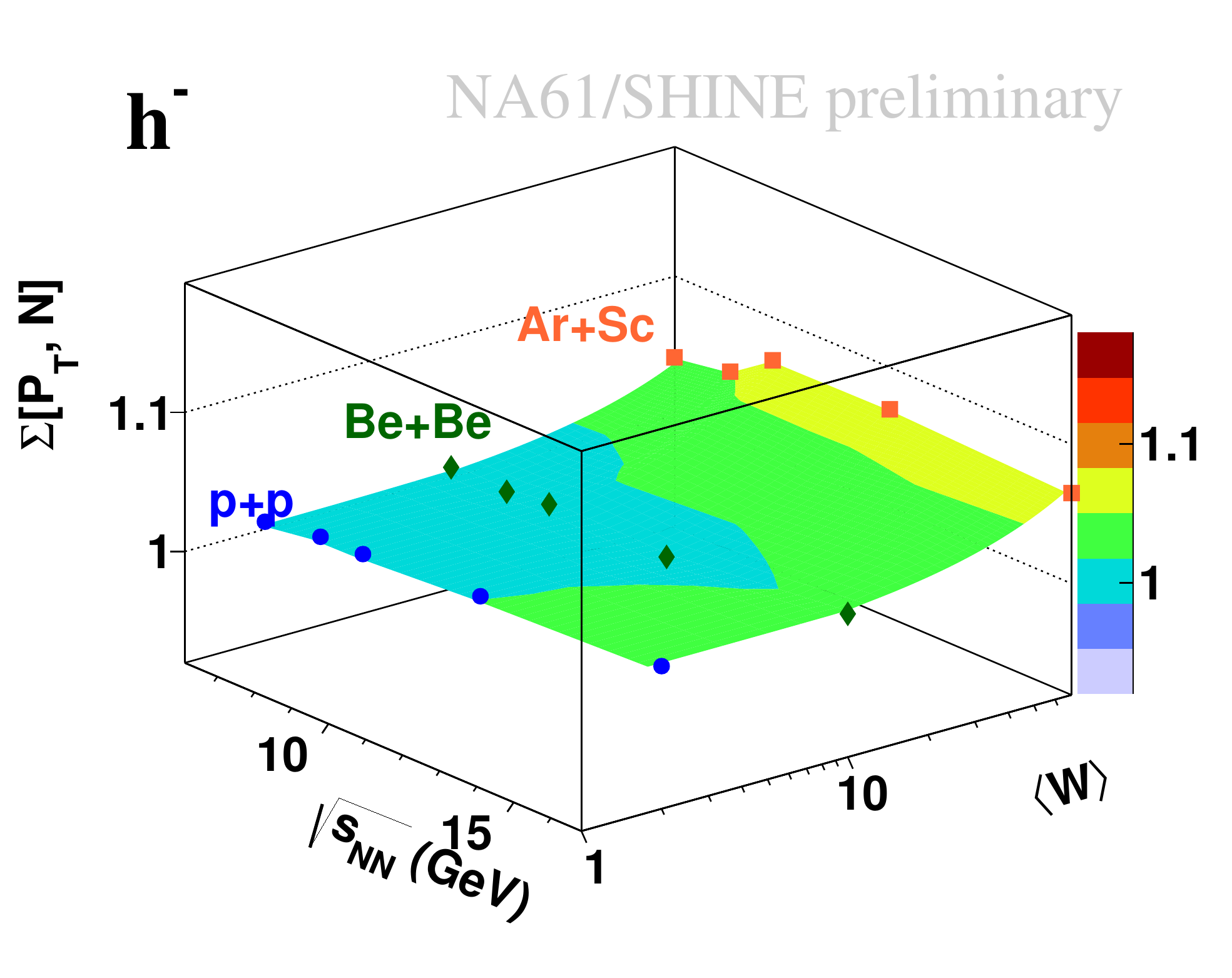}
\vspace{-0.3cm}
\caption[]{\footnotesize Top row: $\Delta[P_T, N]$ and $\Sigma[P_T, N]$ for all charged hadrons ($h^{+}+h^{-}$) in inelastic p+p (grey squares), 0-5\% Be+Be (red triangles), and 0-5\% Ar+Sc (blue circles) collisions obtained by NA61/SHINE in the rapidity interval $0< y_{\pi}< y_{beam}$ and $p_T<1.5$ GeV/c. Only statistical uncertainties are shown. Bottom row: similar as top but results for negatively charged hadrons ($h^{-}$).}
\label{evgeny_pp_bebe_arsc}
\end{figure}

A comparison of preliminary 0-5\% most central Ar+Sc results with preliminary inelastic p+p and 0-5\% central Be+Be collisions is presented in Fig.~\ref{evgeny_pp_bebe_arsc} (note that p+p and Be+Be results were already shown by NA61/SHINE \cite{Aduszkiewicz:2015jna, Czopowicz:2015mfa} but in a slightly different acceptance). 
For all systems values of $\Delta[P_T, N]$ are smaller than 1 and of $\Sigma[P_T, N]$ are higher than 1. A growing deviation of $\Delta[P_T, N]$ and $\Sigma[P_T, N]$ from unity (independent particle model) with energy may be due to increasing azimuthal acceptance (for a given energy azimuthal acceptance for p+p, Be+Be and Ar+Sc is the same). So far, there are no prominent structures in the NA61/SHINE data which could be related to the CP.

\begin{figure}[h]
\centering
\includegraphics[width=0.3\textwidth]{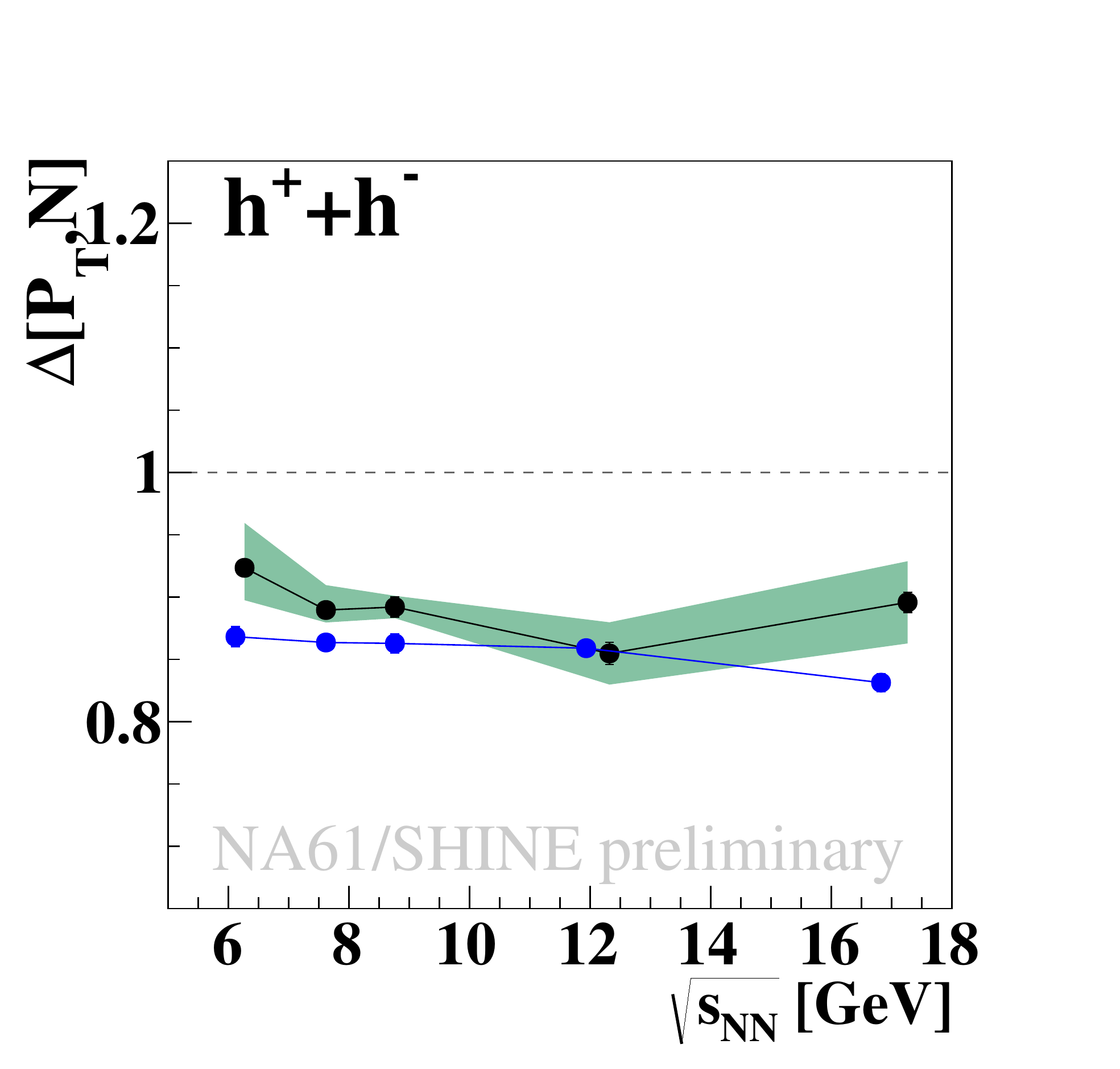}
\includegraphics[width=0.3\textwidth]{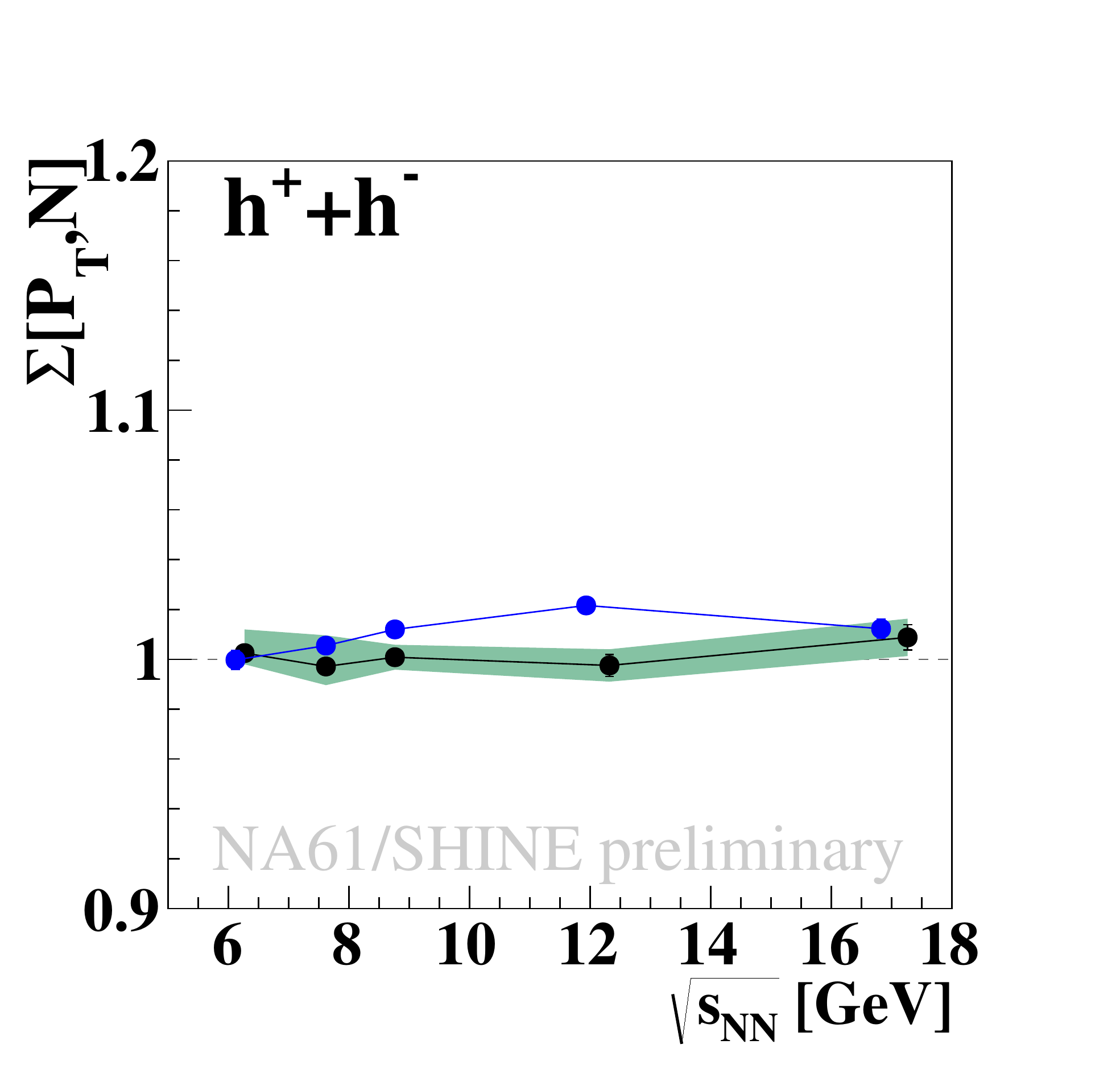}
\vspace{-0.2cm}
\caption[]{\footnotesize $\Delta[P_T, N]$ and $\Sigma[P_T, N]$ in NA61/SHINE 0-5\% Ar+Sc (blue points) and NA49 \cite{Anticic:2015fla} 0-7.2\% central Pb+Pb (black points) collisions. Results for $1.1< y_{\pi}< 2.6$ and $y_{p}< y_{beam}-0.5$ with the same narrow azimuthal angle acceptance for all energies (see Ref.~\cite{Anticic:2015fla} for details). For NA61/SHINE only statistical uncertainties are shown.}
\label{fluct_na49_na61_energy}
\end{figure}

Preliminary NA61/SHINE results have been compared to NA49 data. Figure~\ref{fluct_na49_na61_energy} shows that NA49 Pb+Pb \cite{Anticic:2015fla} and NA61/SHINE Ar+Sc results (in the narrower NA49 acceptance \cite{Anticic:2015fla}) are similar. The values of $\Delta[P_T, N]$ and $\Sigma[P_T, N]$ are now closer to 1 (independent particle model) as expected for a smaller acceptance.

\begin{figure}[h]
\centering
\includegraphics[width=0.3\textwidth]{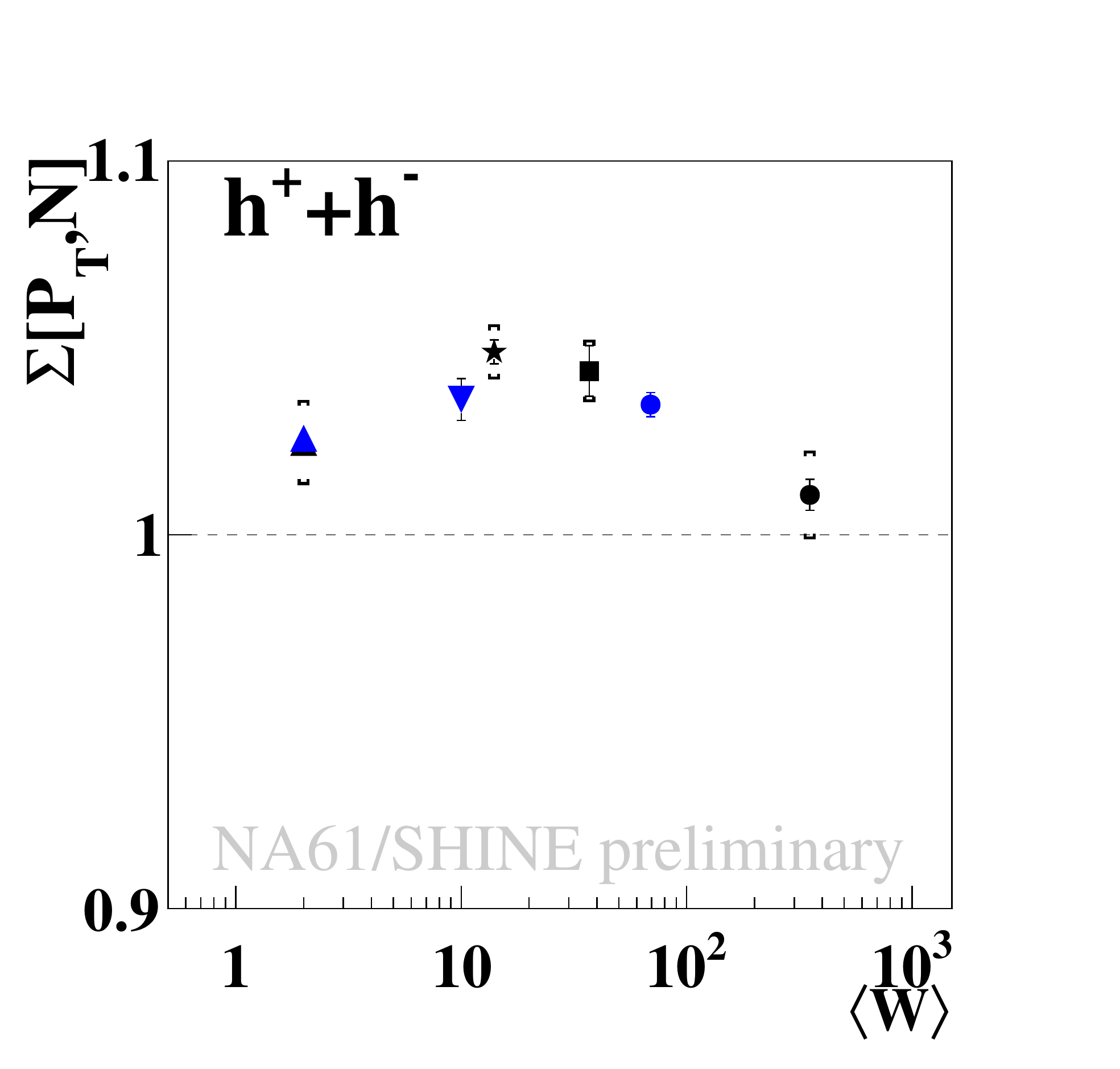}
\includegraphics[width=0.65\textwidth]{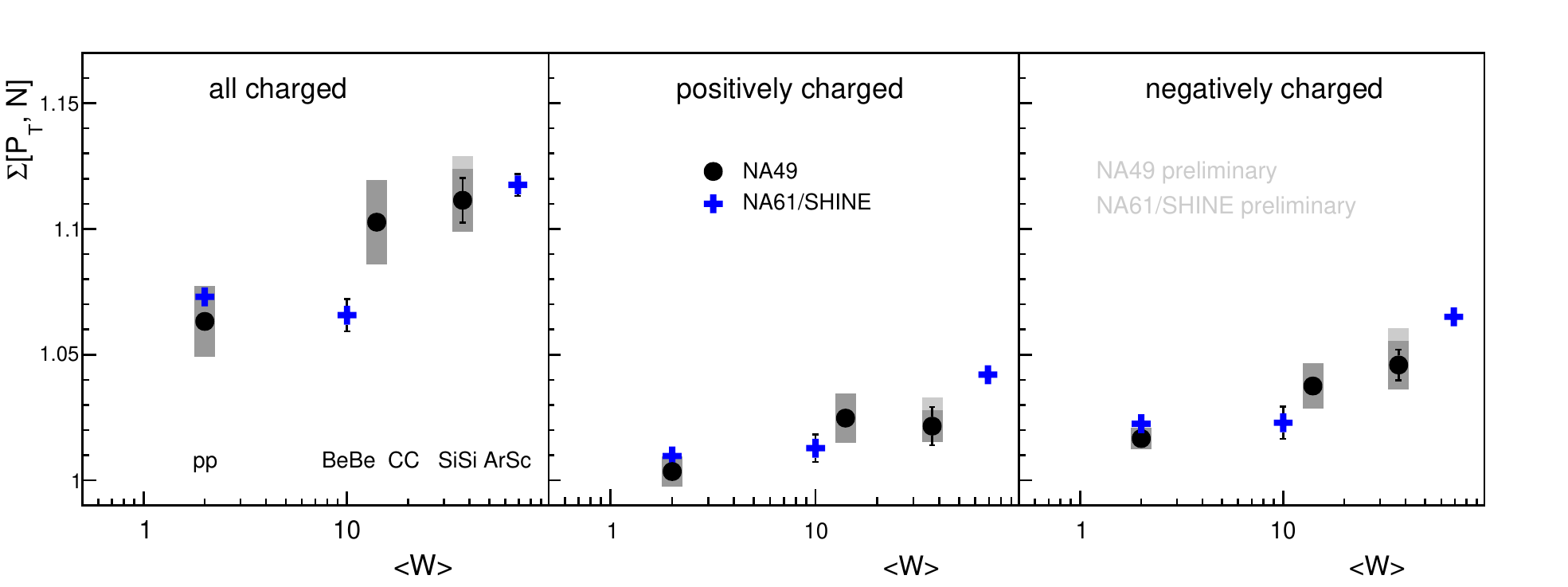}
\vspace{-0.2cm}
\caption[]{\footnotesize Left: $\Sigma[P_T, N]$ at 150/158$A$ GeV/c. Black points are NA49 \cite{Anticic:2015fla} data (p+p, 0-15.3\% C+C, 0-12.2\% Si+Si, 0-5\% Pb+Pb), blue ones preliminary NA61/SHINE (p+p, 0-5\% Be+Be, 0-5\% Ar+Sc). Results in $1.1< y_{\pi}< 2.6$. Right: similar as left (no Pb+Pb point) but results in $0< y_{\pi}< y_{beam}$. Preliminary NA49 \cite{JM_KG_na49_2017} and NA61/SHINE data. For NA61/SHINE only statistical uncertainties are shown.}
\label{fluct_na49_na61_size}
\end{figure}

For the system size dependence of $\Sigma[P_T, N]$ at 150/158$A$ GeV/c (Fig.~\ref{fluct_na49_na61_size}, left) the NA49 \cite{Anticic:2015fla} and NA61/SHINE data points show consistent trends. $\Delta[P_T, N]$ (not shown) is more sensitive to the width of the centrality interval \cite{Gorenstein:2013nea} and points are scattered (see Ref.~\cite{evgeny_cpod16_slides}).
So far the analysis in NA49 was restricted to $1.1< y_{\pi}< 2.6$ but in 2017 p+p, C+C and Si+Si data were reanalysed in the acceptance currently used by NA61/SHINE ($0< y_{\pi}< y_{beam}$). As expected, fluctuations are larger for the larger rapidity interval (Fig.~\ref{fluct_na49_na61_size}, right) and an increase of $\Sigma[P_T, N]$ from p+p to Ar+Sc/Si+Si can be observed. It is now very important to obtain future NA61/SHINE results in Xe+La and Pb+Pb to figure out whether a monotonic or rather non-monotonic (as seen in Fig.~\ref{fluct_na49_na61_size}, left) dependence will be observed in this larger rapidity window.

\section{Multiplicity fluctuations in p+p, Be+Be and Ar+Sc}

Multiplicity fluctuations can be determined using the scaled variance, $\omega[N]$, of the multiplicity distribution (see Ref.~\cite{Aduszkiewicz:2015jna}). In case of no fluctuations $\omega[N]$ equals 0, and it is equal to 1 for the Poisson multiplicity distribution. The quantity $\omega[N]$ is an {\it intensive} but not a {\it strongly intensive} measure of fluctuations, thus in the Grand Canonical Ensemble it does not depend on the volume but depends on volume fluctuations. In order to reduce this effect $\omega[N]$ is shown here for very central collisions, namely 0-0.2\% Ar+Sc and 0-1\% Be+Be. The NA61/SHINE acceptance used in this analysis is the same as for transverse momentum and multiplicity fluctuations (previous section).

\begin{figure}[h]
\centering
\includegraphics[width=0.3\textwidth]{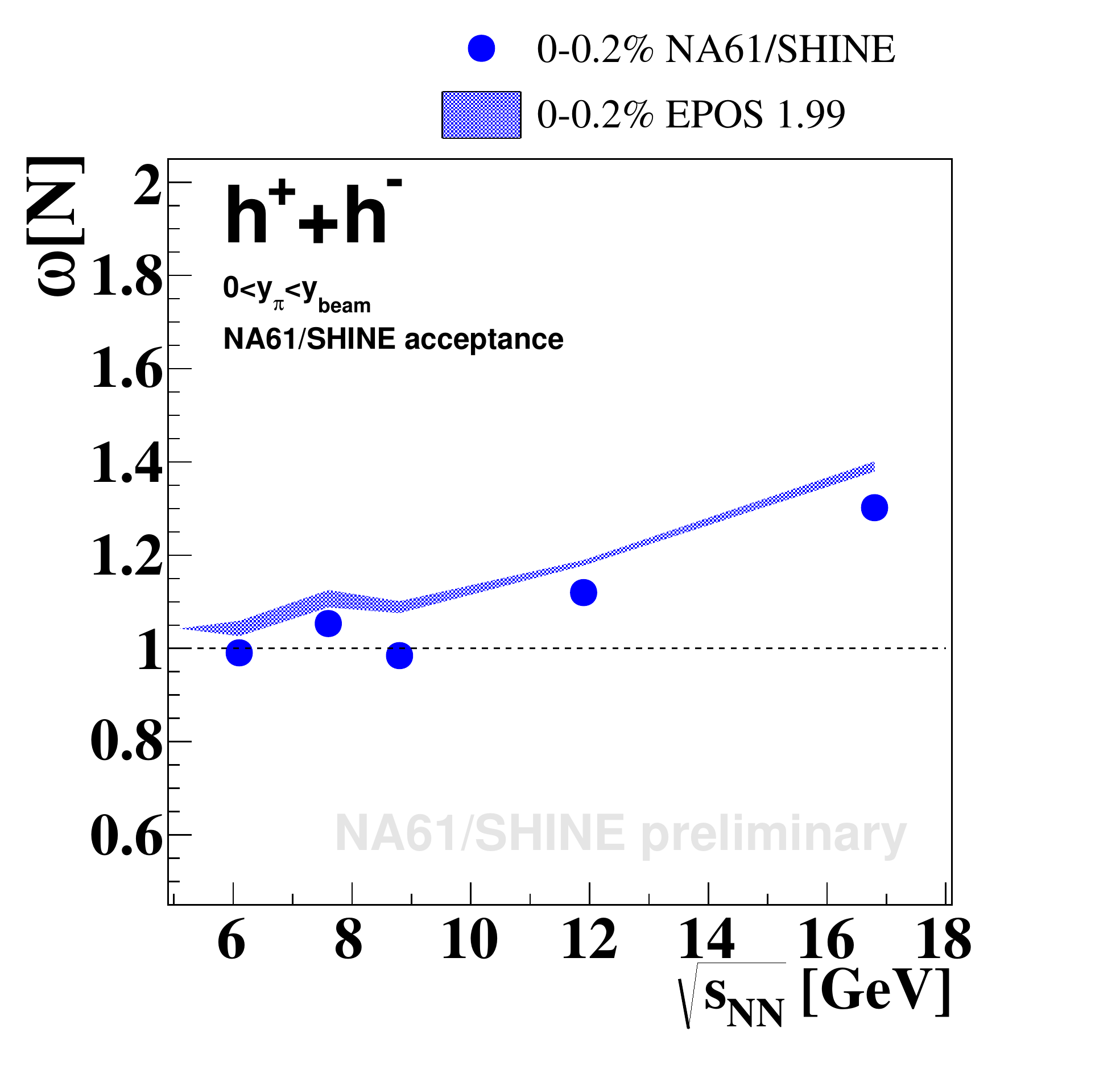}
\includegraphics[width=0.3\textwidth]{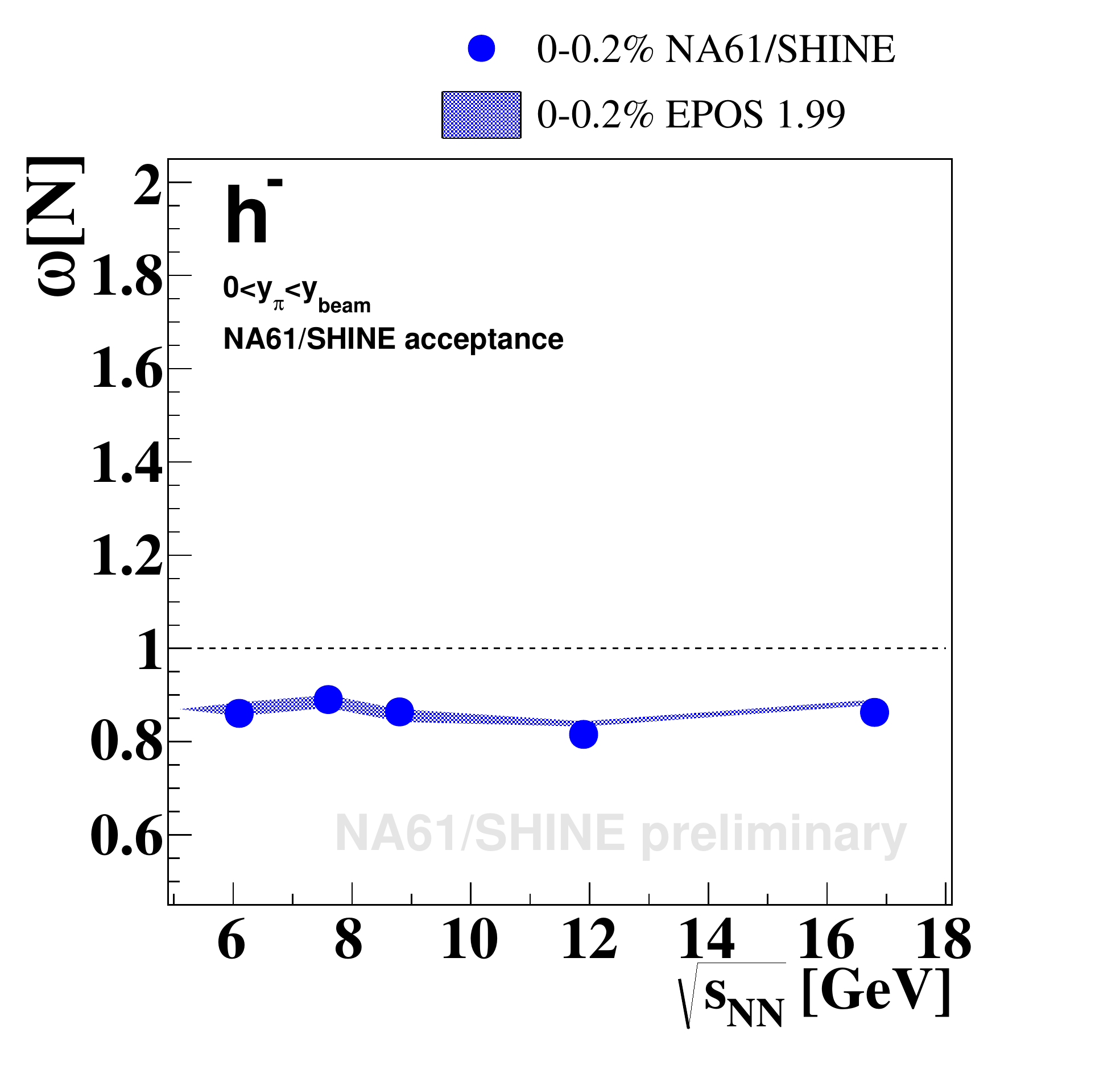}
\includegraphics[width=0.3\textwidth]{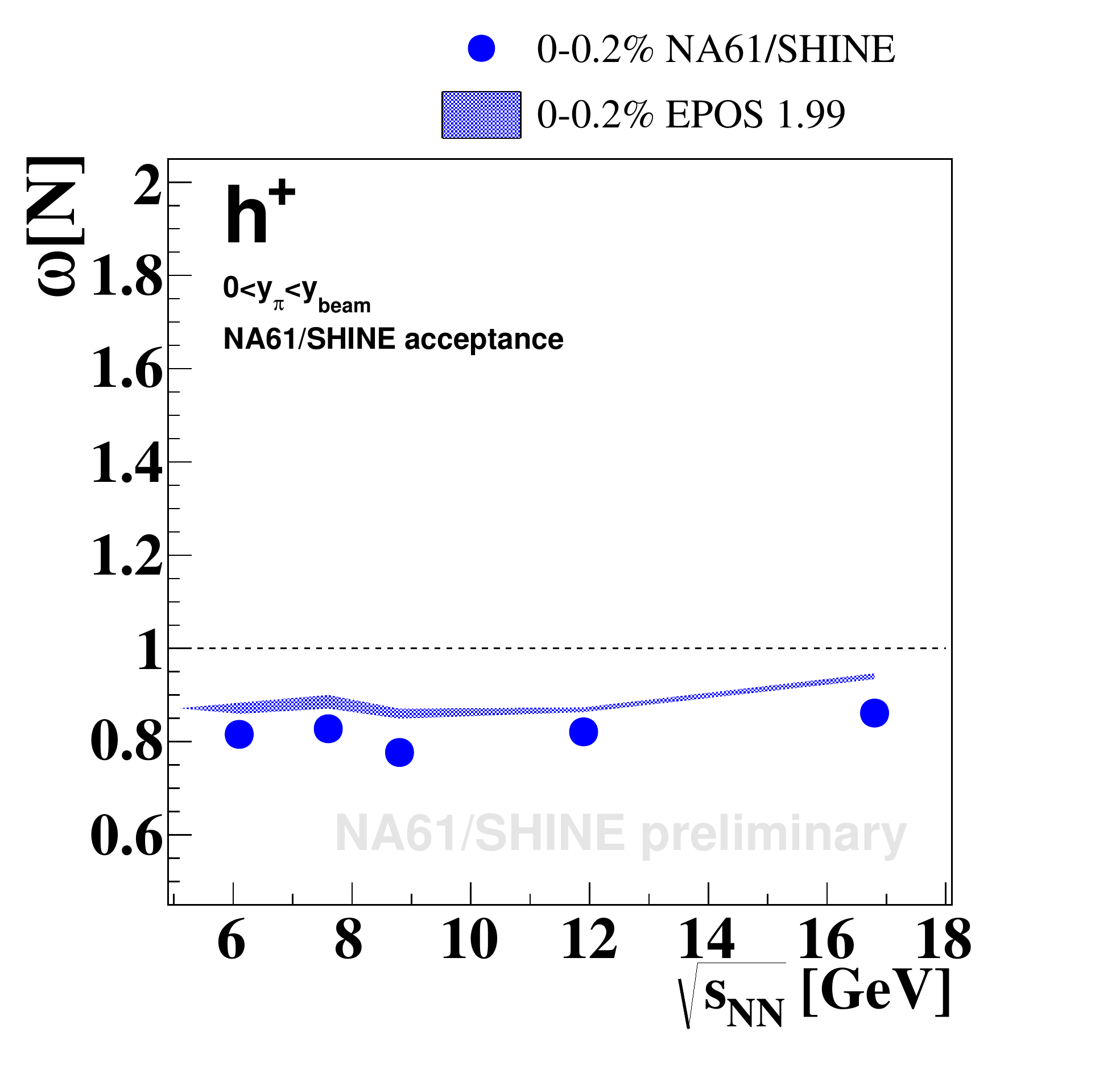}
\vspace{-0.2cm}
\caption[]{\footnotesize $\omega[N]$ measure of multiplicity fluctuations obtained by NA61/SHINE for 0-0.2\% central Ar+Sc collisions in $0< y_{\pi}< y_{beam}$ and $p_T<1.5$ GeV/c for all charged ($h^{+}+h^{-}$), negatively charged ($h^{-}$) and positively charged ($h^{+}$) hadrons. Only statistical uncertainties are shown. NA61/SHINE data are compared to EPOS 1.99.}
\label{andrey_arsc_epos}
\end{figure}

Figure \ref{andrey_arsc_epos} shows that there is no significant non-monotonic behavior in the energy dependence of $\omega[N]$. An increase of $\omega[N]$ (for charged hadrons) with energy may reflect the increase of $\omega[N_{ch}]$ measured in p+p collisions in full phase-space (KNO scaling, see Ref.~\cite{Heiselberg:2000fk}). The NA61/SHINE results for 0-0.2\% Ar+Sc are in agreement with the EPOS 1.99 model.

\begin{wrapfigure}{r}{9cm}
\centering
\vspace{-0.4cm}
\includegraphics[width=0.35\textwidth]{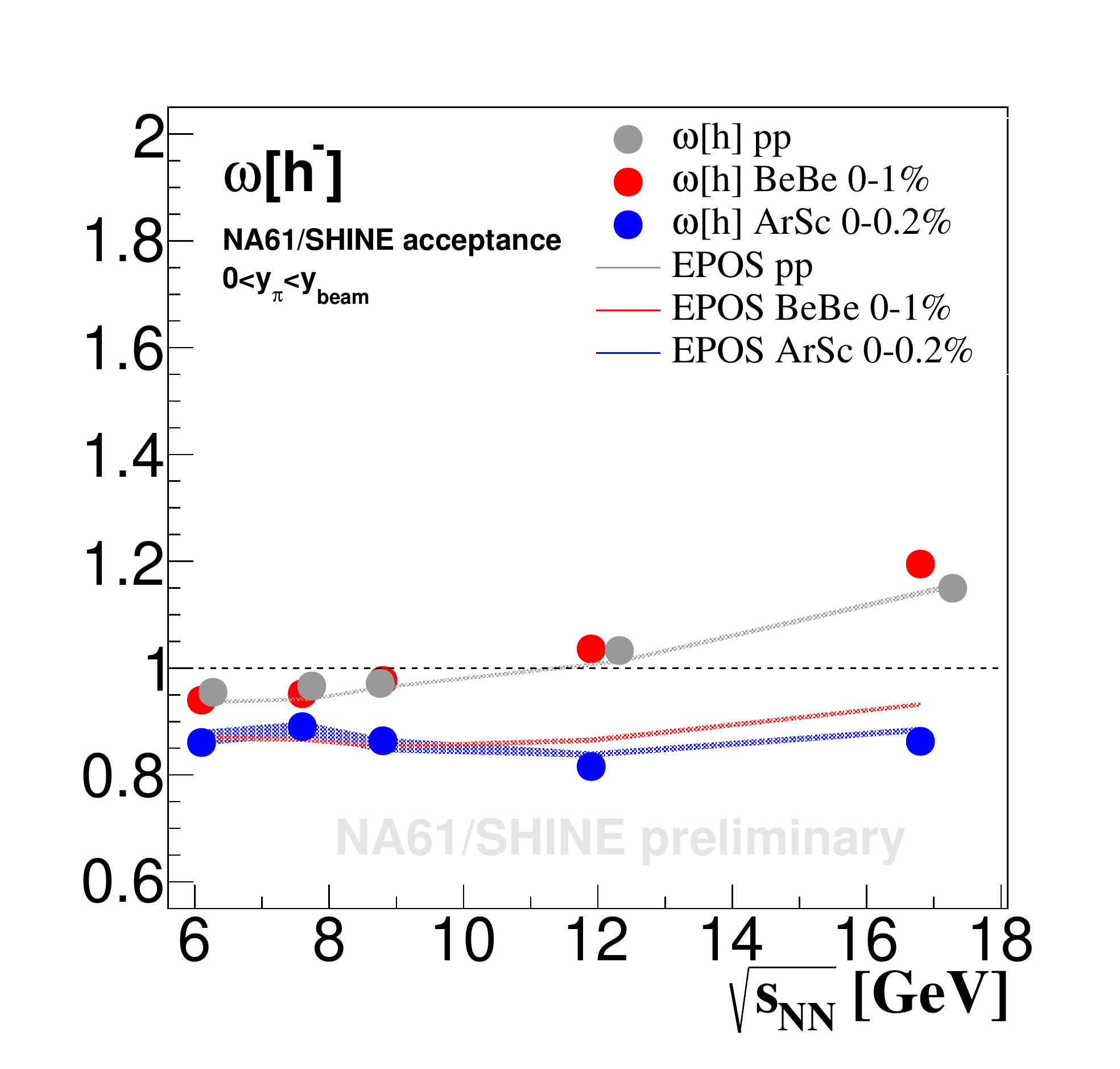}
\vspace{-0.25cm}
\caption[]{\footnotesize $\omega[N]$ in inelastic p+p (grey), 0-1\% Be+Be (red), and 0-0.2\% Ar+Sc (blue) collisions obtained by NA61/SHINE in $0< y_{\pi}< y_{beam}$ and $p_T<1.5$ GeV/c. Results for negatively charged hadrons. Only statistical uncertainties are shown. NA61/SHINE data are compared to EPOS 1.99.}
\label{andrey_pp_bebe_arsc_epos}
\end{wrapfigure}

A comparison of multiplicity fluctuations for p+p, 0-1\% Be+Be and 0-0.2\% Ar+Sc is shown in Fig.~\ref{andrey_pp_bebe_arsc_epos}. For all studied systems there is no non-monotonic dependence of multiplicity fluctuations with energy. However, a different energy dependence is seen for Be+Be (and p+p) and Ar+Sc collisions! The EPOS model reproduces Ar+Sc and p+p results and underestimates Be+Be data.

\begin{figure}[h]
\centering
\includegraphics[width=0.3\textwidth]{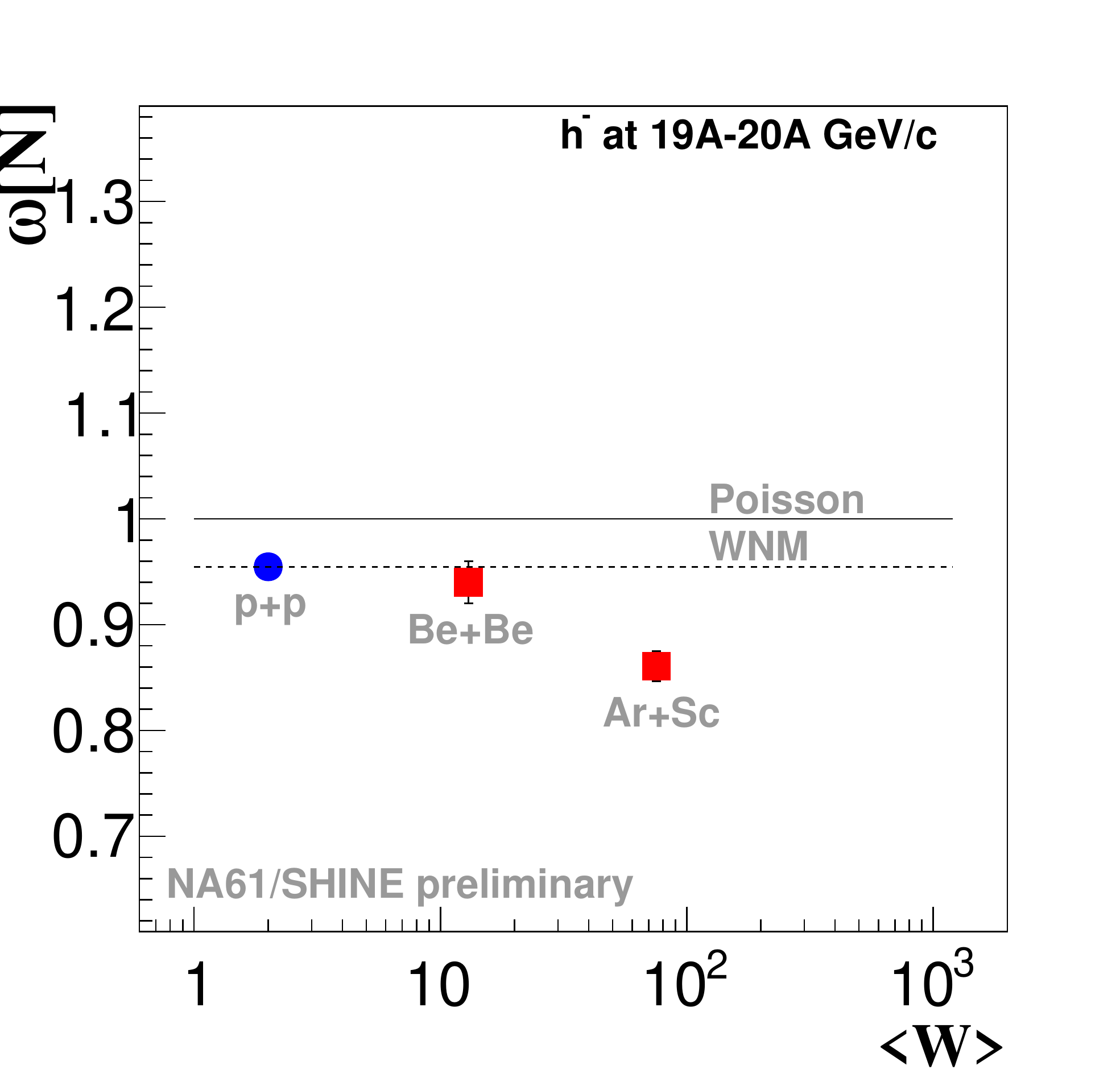}
\includegraphics[width=0.3\textwidth]{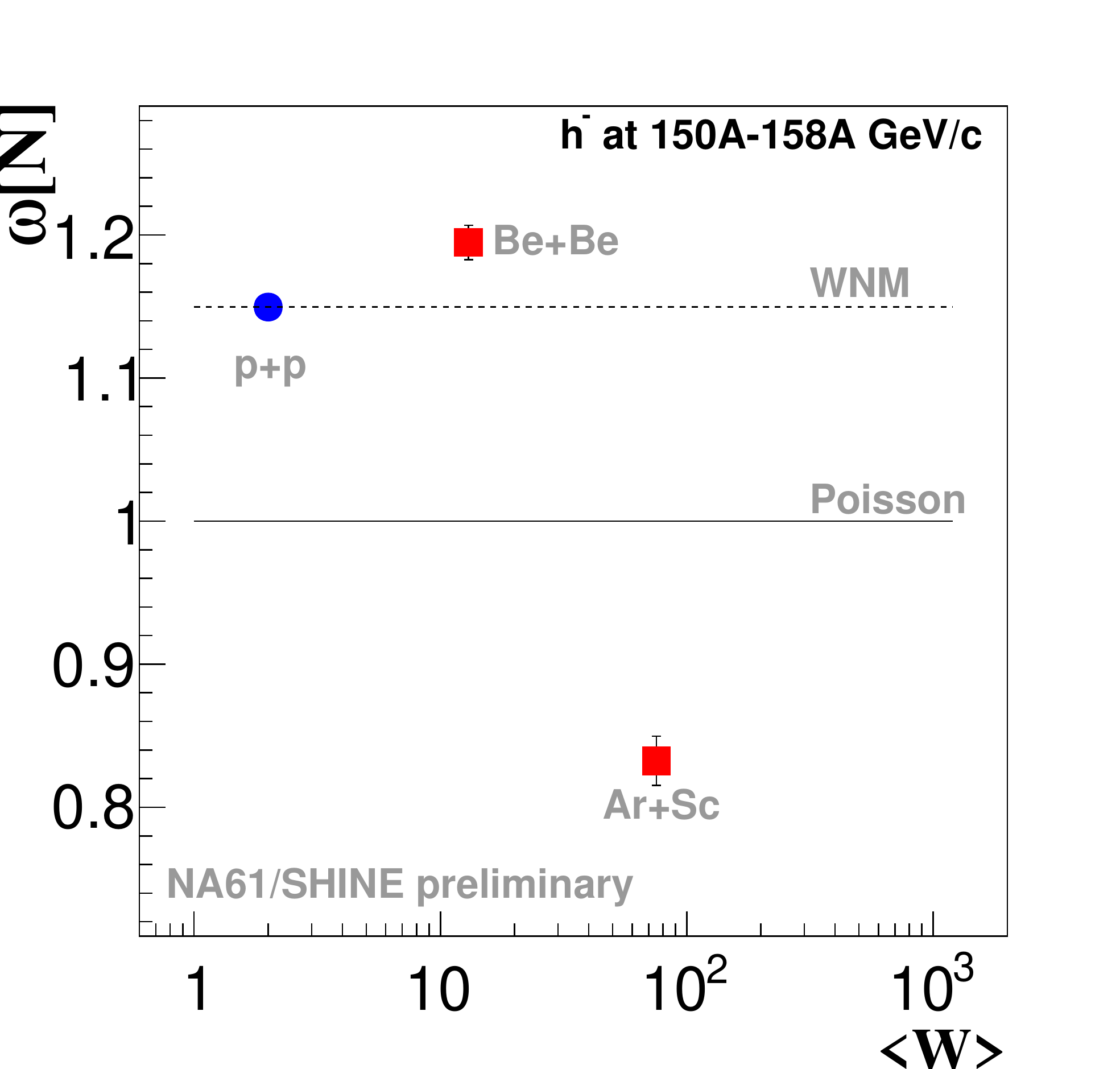}
\vspace{-0.22cm}
\caption[]{\footnotesize Multiplicity fluctuations of negatively charged hadrons versus mean number of wounded nucleons at low (left) and high (right) SPS energy. Results for inelastic p+p, 0-1\% Be+Be, and 0-0.2\% central Ar+Sc collisions are presented for $0< y_{\pi}< y_{beam}$ and $p_T<1.5$ GeV/c. Only statistical uncertainties are shown.}
\label{omega_Nw}
\end{figure}

Figures \ref{andrey_pp_bebe_arsc_epos} and \ref{omega_Nw} show an interesting effect. Namely, for negatively charged hadrons (the cleanest sample) multiplicity fluctuations are larger in p+p and central Be+Be than in central Ar+Sc. Within Wounded Nucleon Model without $W$ fluctuations $\omega[N]_{AA}=\omega[N]_{pp}$ (dotted lines in Fig.~\ref{omega_Nw}); WNM with $W$ fluctuations results in $\omega[N]_{AA}>\omega[N]_{pp}$. Thus Ar+Sc results show violation of the Wounded Nucleon Model in the fluctuation analysis. In the Ideal Boltzmann Grand Canonical Ensemble (IB-GCE) $\omega[N]=1$ (Poisson) independently of the (fixed) system volume. Thus $\omega[N]_{AA}<1$ is forbidden in the IB-GCE. However, $\omega[N]$ can increase due to resonance decays and Bose-Einstein correlation effects. On the other hand, a decrease of the values of $\omega[N]$ (as seen in Ar+Sc) may be due to conservation laws \cite{Begun:2006uu}. Finally, $\omega[N] \gg 1$, as seen in p+p at 158$A$ GeV/c, can be understood in statistical models as a result of volume and/or energy fluctuations~\cite{Begun:2008fm}.

\section{Higher order moments of net-charge and multiplicity distributions in p+p}

Higher order moments of multiplicity distributions (skewness $S$, kurtosis~$\kappa$) measure the non-Gaussian nature of fluctuations and are more sensitive (than the variance $\sigma^2$) to fluctuations at a CP \cite{Stephanov:2008qz, Stephanov:2011pb}. 
Moreover, they can be used to test (statistical and dynamical) models (first moments do not allow to distinguish between different types of models; already for second moments fluctuations are different in string and statistical models). 
Finally, higher moments of conserved quantum numbers ($i = B, Q, S$) allow for direct comparison to theory via susceptibilities ($S\sigma \approx \chi_i^3 / \chi_i^2$, $\kappa \sigma^2 \approx \chi_i^4 / \chi_i^2$). It is worth mentioning that $\omega$, as well as products of higher order moments, $S \sigma$ and $\kappa \sigma^2$, are {\it intensive} measures of fluctuations.

\begin{figure}[h]
\centering
\includegraphics[width=0.32\textwidth]{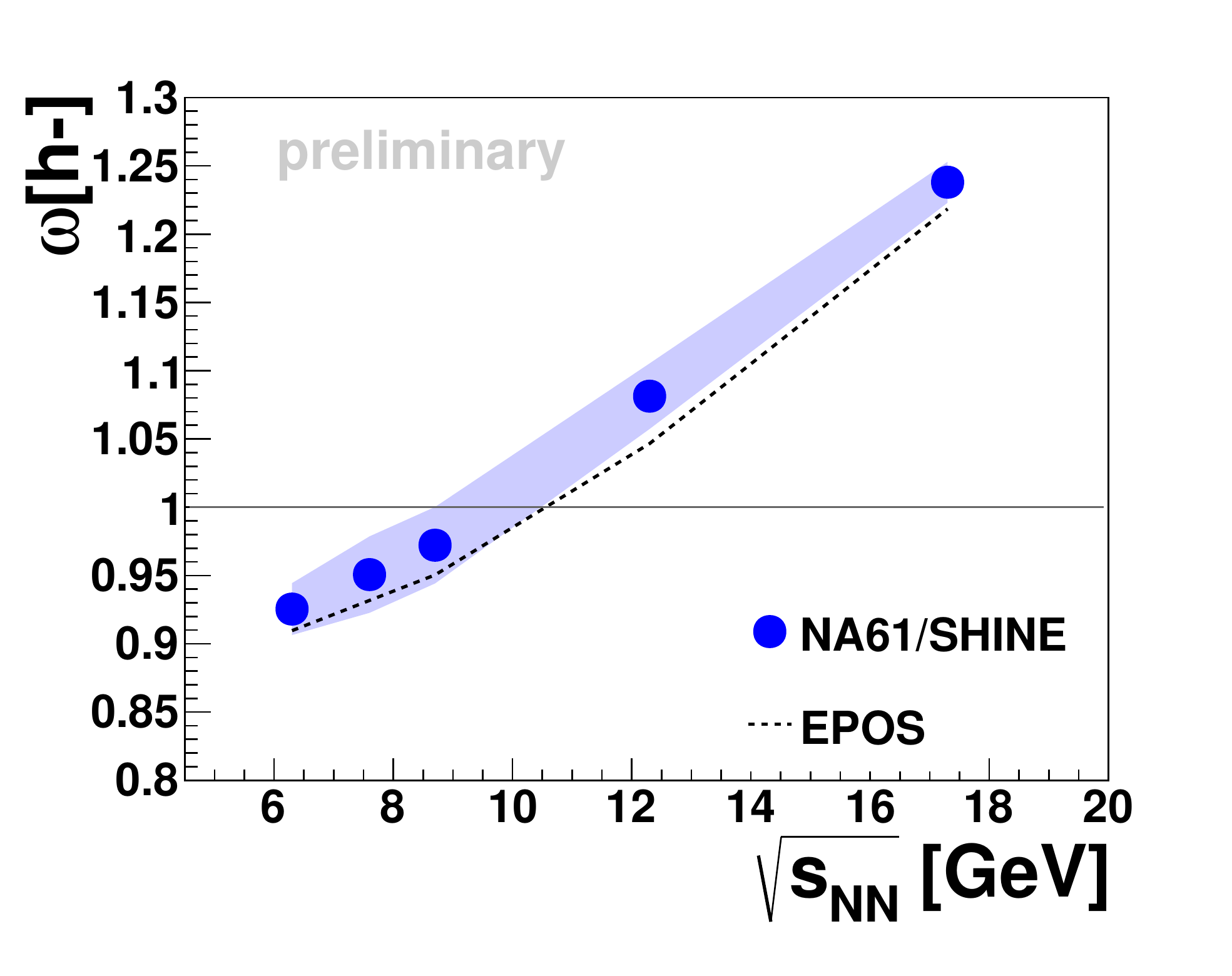}
\includegraphics[width=0.32\textwidth]{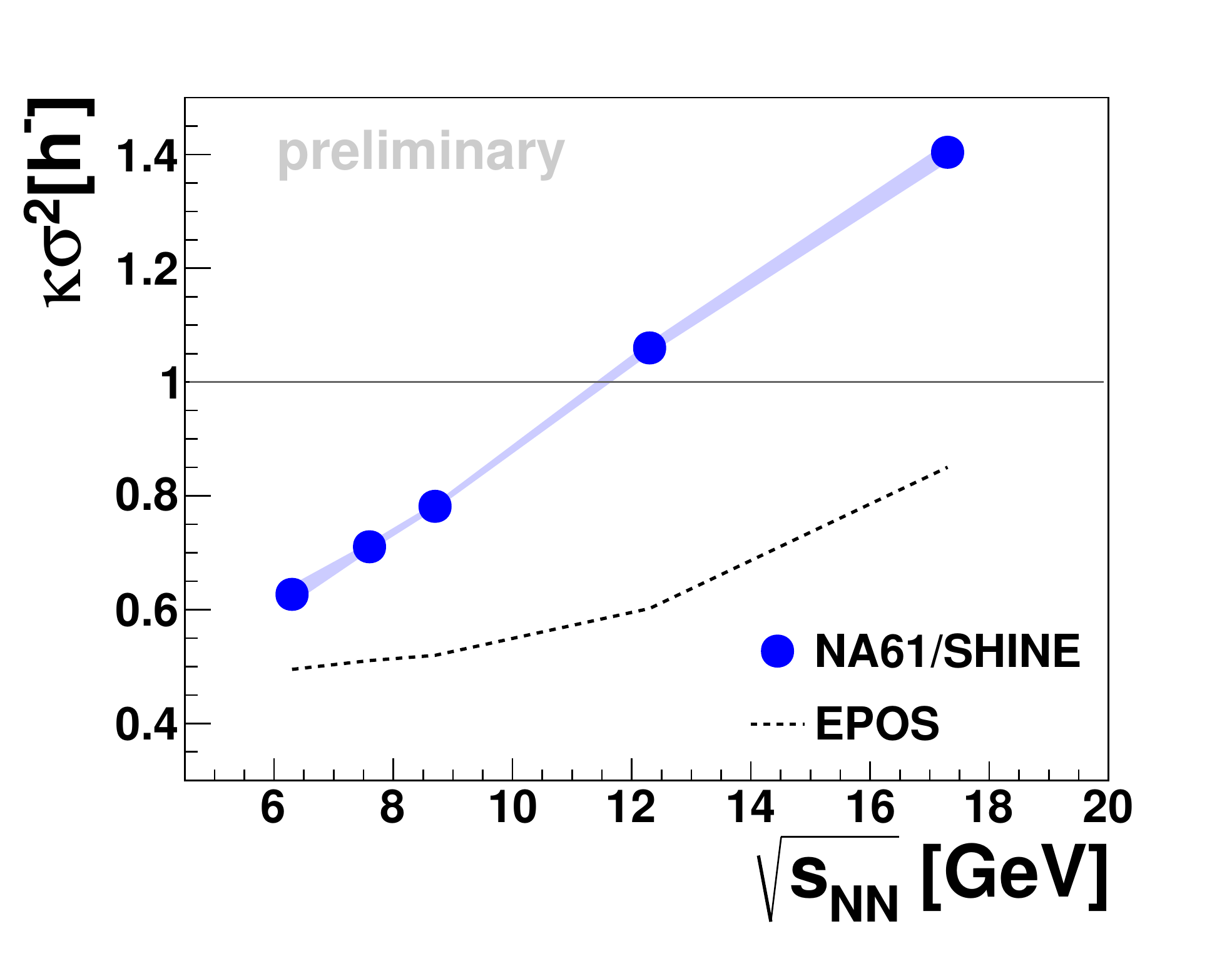}
\includegraphics[width=0.32\textwidth]{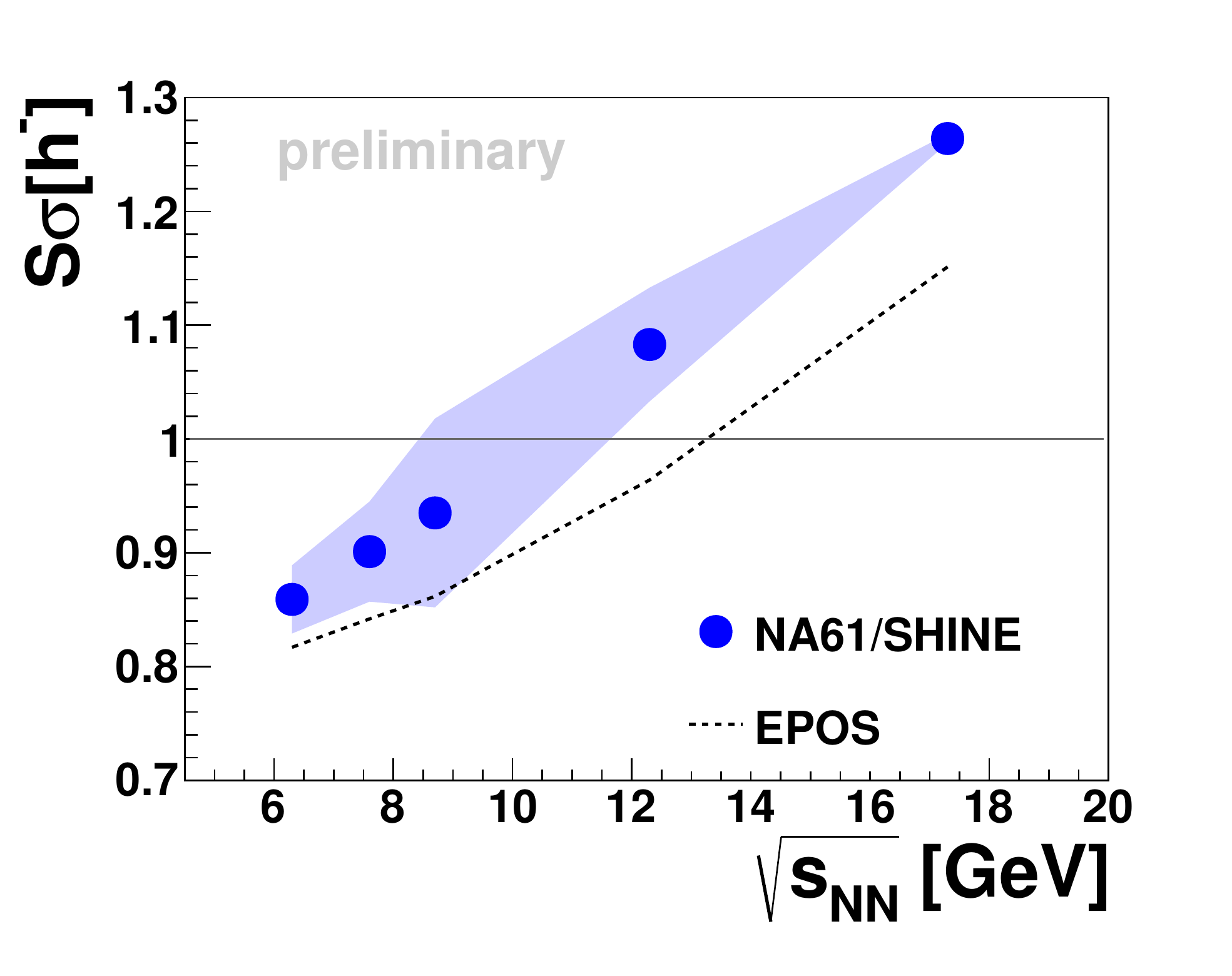}
\vspace{-0.3cm}
\caption[]{\footnotesize Scaled variance, $\omega$, and products of higher order moments of multiplicity distribution of negatively charged hadrons measured in inelastic p+p interactions in acceptance as defined in Ref.~\cite{Aduszkiewicz:2015jna} (without additional rapidity cut).}
\label{mmp_hminus}
\end{figure}

\begin{figure}[ht]
\centering
\includegraphics[width=0.32\textwidth]{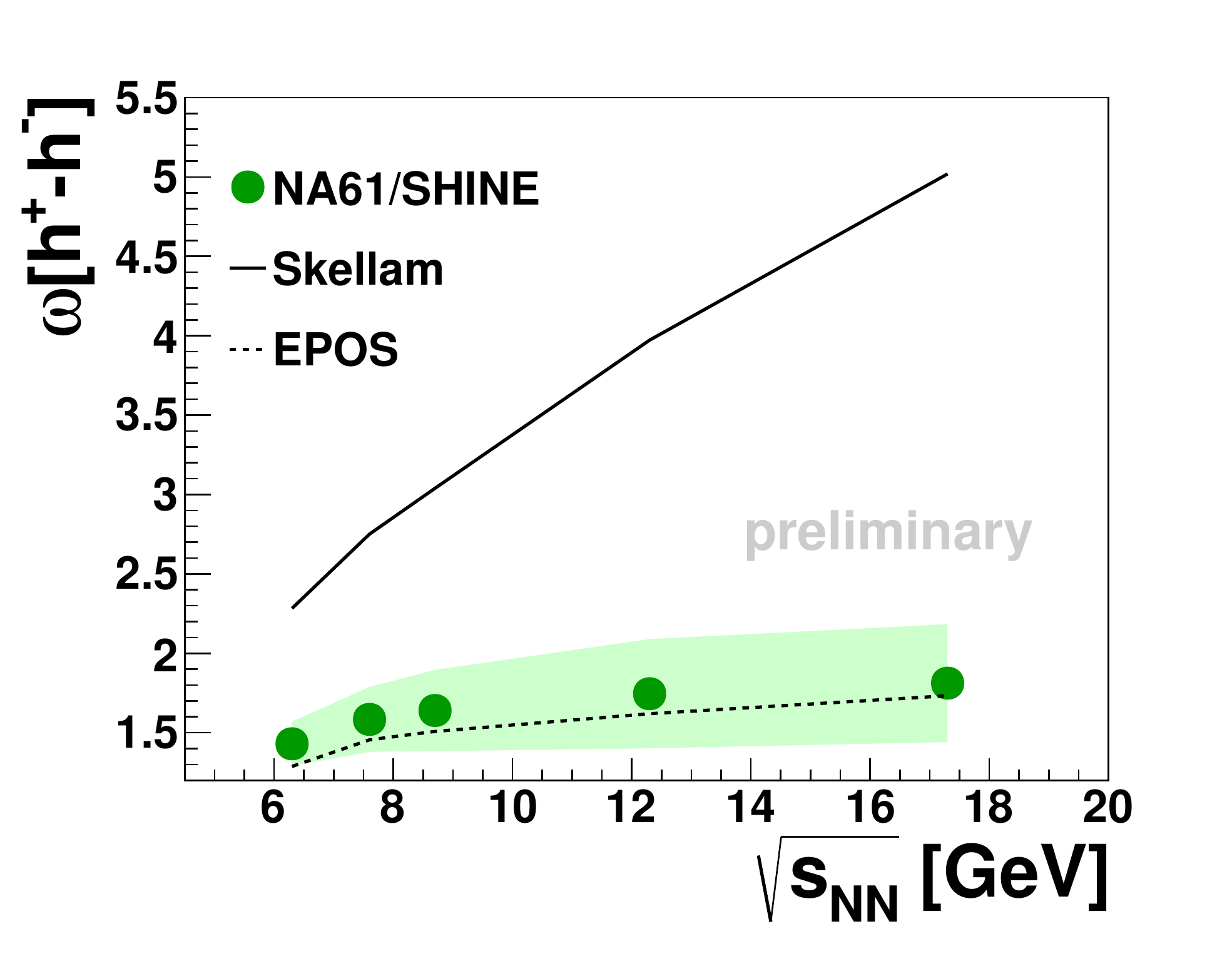}
\includegraphics[width=0.32\textwidth]{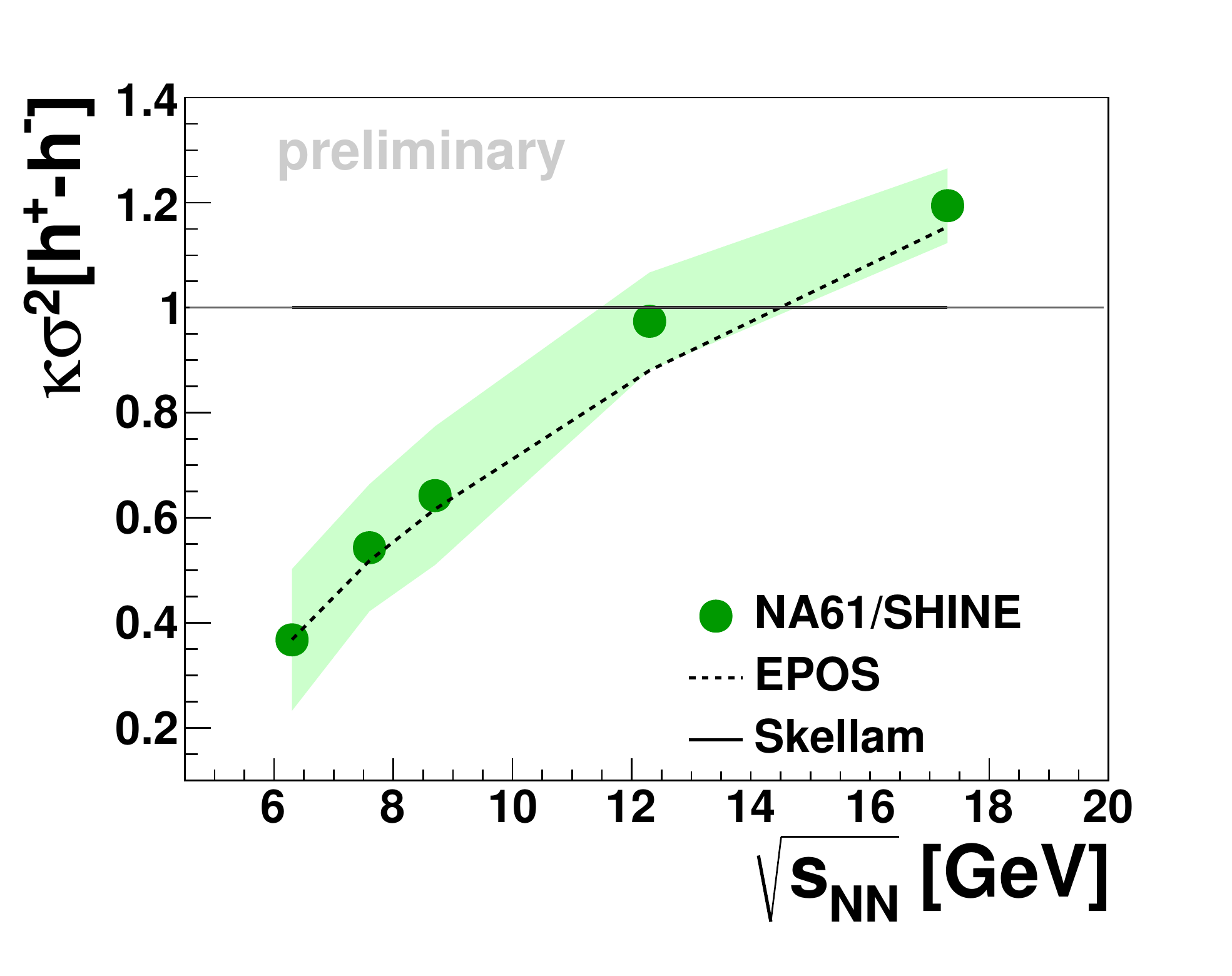}
\includegraphics[width=0.32\textwidth]{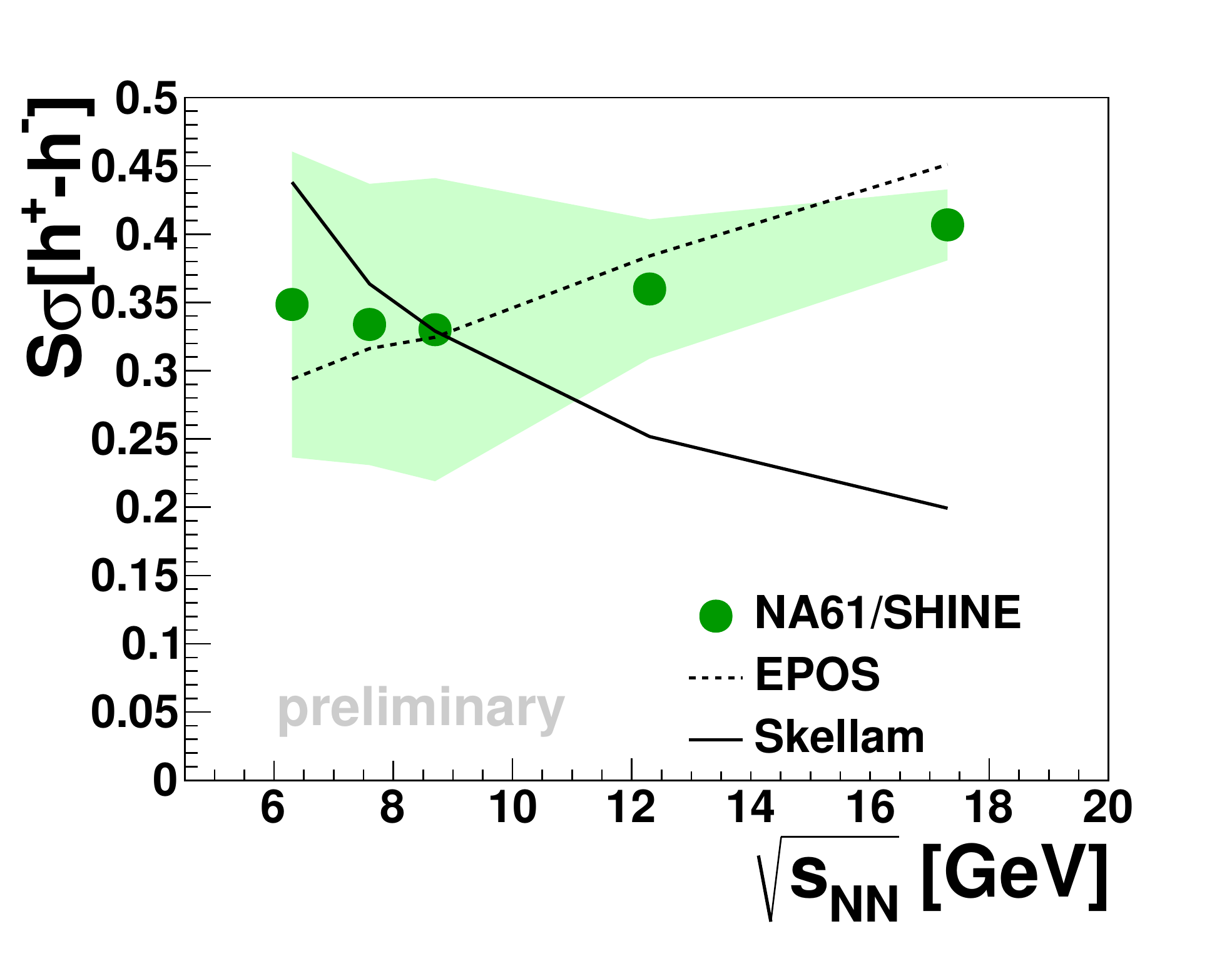}
\vspace{-0.3cm}
\caption[]{\footnotesize Scaled variance, $\omega$, and products of higher order moments of net-charge distributions measured in inelastic p+p interactions in acceptance as defined in Ref.~\cite{Aduszkiewicz:2015jna} (without additional rapidity cut). }
\label{mmp_net_charge}
\end{figure}

The scaled variance, $\omega$, and products of higher order moments of multiplicity distributions of negatively charged hadrons as well as of net-charge distributions measured in inelastic p+p interactions are shown in Figs.~\ref{mmp_hminus} and \ref{mmp_net_charge}. There is no non-monotonic behavior suggesting a CP. Tendencies of all measures are reproduced by the EPOS 1.99 model but the magnitude of $\kappa \sigma^2[h^{-}]$ is not reproduced. Results on net-charge fluctuations do not agree with independent particle production (Skellam distribution). The collisions of Be+Be and Ar+Sc are planned to be analyzed soon.

\section{Summary}

In this report preliminary NA61/SHINE results on transverse momentum and multiplicity fluctuations in p+p, Be+Be and Ar+Sc collisions were presented for the rapidity interval $0 < y_{\pi} < y_{beam}$. So far the energy dependences did not show non-monotonic behavior as might be expected for a CP. We are waiting for Xe+La and Pb+Pb data. 

Preliminary results on multiplicity fluctuations in p+p, Be+Be and Ar+Sc interactions were calculated in the same rapidity range. The energy dependences also do not show non-monotonic behavior as expected for a CP. Different energy dependences are seen for Be+Be and Ar+Sc collisions. Fluctuations in Ar+Sc are smaller than in p+p. Thus the Ar+Sc results cannot be explained within the Wounded Nucleon Model.

Results on higher order moments of net-charge and multiplicity distributions in p+p collisions were shown. The trends of all measures are reproduced by EPOS but results on net-charge fluctuations do not agree with independent particle production.

There are many ongoing analyses in NA61/SHINE such as proton intermittency in Be+Be and Ar+Sc, two-particle correlations in azimuthal angle and pseudorapidity in Be+Be (see NA61/SHINE talks at CPOD 2017), higher order moments of net-charge in Be+Be, etc. The Pb+Pb data taking started in 2016 and the full energy scan of Xe+La is scheduled for this year. Stay tuned!



\begin{thebibliography}{99}

\bibitem{Aduszkiewicz:2015jna} 
  A.~Aduszkiewicz {\it et al.} [NA61/SHINE Collaboration],
  Eur.\ Phys.\ J.\ C {\bf 76}, no. 11, 635 (2016)



\bibitem{Gorenstein:2013nea} 
  M.~I.~Gorenstein and K.~Grebieszkow,
  Phys.\ Rev.\ C {\bf 89}, no. 3, 034903 (2014)



\bibitem{Czopowicz:2015mfa} 
  T.~Czopowicz [NA61/SHINE Collaboration],
  PoS CPOD {\bf 2014}, 054 (2015)



\bibitem{Anticic:2015fla} 
  T.~Anticic {\it et al.} [NA49 Collaboration],
  Phys.\ Rev.\ C {\bf 92}, no. 4, 044905 (2015)



\bibitem{evgeny_cpod16_slides} E.~Andronov, slides from Critical Point and Onset of Deconfinement 2016, 

https://indico.cern.ch/event/449173/contributions/2167165/attachments/1280888/1902801/EA\_cpod.pdf


\bibitem{JM_KG_na49_2017} J.~Miernik, K.~Grebieszkow, NA49 preliminary results (2017)



\bibitem{Heiselberg:2000fk} 
  H.~Heiselberg,
  Phys.\ Rept.\  {\bf 351}, 161 (2001)



\bibitem{Begun:2006uu} 
  V.~V.~Begun, M.~Gazdzicki, M.~I.~Gorenstein, M.~Hauer, V.~P.~Konchakovski and B.~Lungwitz, \\
  Phys.\ Rev.\ C {\bf 76}, 024902 (2007)



\bibitem{Begun:2008fm} 
  V.~V.~Begun, M.~Gazdzicki and M.~I.~Gorenstein,
  Phys.\ Rev.\ C {\bf 78}, 024904 (2008)



\bibitem{Stephanov:2008qz} 
  M.~A.~Stephanov,
  Phys.\ Rev.\ Lett.\  {\bf 102}, 032301 (2009)



\bibitem{Stephanov:2011pb} 
  M.~A.~Stephanov,
  Phys.\ Rev.\ Lett.\  {\bf 107}, 052301 (2011)




\end{thebibliography}
\end{document}